\documentclass[preprint,preprintnumbers,amsmath,amssymb]{revtex4-1}

\usepackage{graphicx}
\usepackage{epsfig}
\usepackage{bm}

\def\be{\begin{equation}}
\def\ee{\end{equation}}
\def\bee{\begin{eqnarray}}
\def\eee{\end{eqnarray}}

\begin{document}

\preprint{Preprint: do not distribute}

\title{Nonlinear seed island generation by three-dimensional electromagnetic, gyrokinetic turbulence.}
\author{W.~A.~Hornsby, P.~Migliano, R.~Buchholz, A.G.~Peeters}
\affiliation{Theoretical Physics V, Dept. of Physics, Universitaet Bayreuth, Bayreuth, Germany, D-95447}
\email{william.hornsby@uni-bayreuth.de}

\author{D.~Zarzoso, F.J.~Casson, E.~Poli}
\affiliation{Max-Planck-Institut f\" ur Plasmaphysik, Boltzmannstrasse 2, D-85748
Garching bei M\" unchen, Germany} 

\date{\today}

\begin{abstract}

Turbulence is shown to be critical to the onset and evolution of the neoclassical tearing mode, affecting both its growth and rotation.  The interaction is here studied for the first time in the three dimensional, toroidal gyrokinetic framework.   Turbulent fluctuations do not destroy
the growing island early in its development, which maintains a coherent form as it grows, in fact the island is seeded and its rotation frequency determined, by nonlinear interaction.  This process provides an initial structure that is of the order of an ion gyro-radius 
wide, allowing the island to rapidly reach a large size.  A large degree of stochastisation around the seperatrix, and a
complete breakdown of the X-point is seen, which significantly reduces the effective island width.  A turbulent modification of the electrostatic field in and around the island greatly affects the size of the resonant layer width, and the island is seen to grow at the linear rate even though the island is 
significantly wider than the singular layer width.  

\end{abstract}

\pacs{}
\keywords{Turbulence, plasma, multiscale}
\maketitle

Magnetic islands in a tokamak can lead to loss of confinement through a change of magnetic topology via magnetic reconnection.  Their generation can lead to major disruptions of 
confined plasmas. As a matter of fact the tearing mode \cite{FUR63,FUR73} and specifically the neoclassical tearing mode (NTM) \cite{CAR86,WIL96}, is 
expected to set the beta limit in a reactor \cite{WAE09}.  On the other hand, drift-wave turbulence is widely acknowledged to be cause of
anomalous transport which, in turn, is regulated by zonal and mesoscale flows \cite{ter00}.  The generation of large scale structures by the turbulence has 
been suggested as a possible mechanism for the generation of seed islands, vital to the evolution and stability
of the NTM \cite{itoh04,ish10}.

Plasma turbulence and tearing modes occupy disparate time and length scales, with 
turbulence occupying the micro-scale defined by the ion gyro-radius and drift frequency.  Tearing modes occupy a significant fraction
of a toroidal turn, however, early in their evolution, islands can be very narrow and thus comparable
to turbulent length scales.  As such, their evolution can not be considered to be independent of the turbulence \cite{MIL09}.
The radial extent of the magnetic island evolves over a resistive time scale, which, in high temperature, weakly collisional fusion plasmas is longer than the turbulence time scales.   
Due to the complexity of the problem, analytical 
theory in this field is difficult.  Here, we approach the problem using massively parallel, 
state-of-the-art kinetic simulation.  

\begin{figure}
\centering
\includegraphics[width=8.5cm,clip]{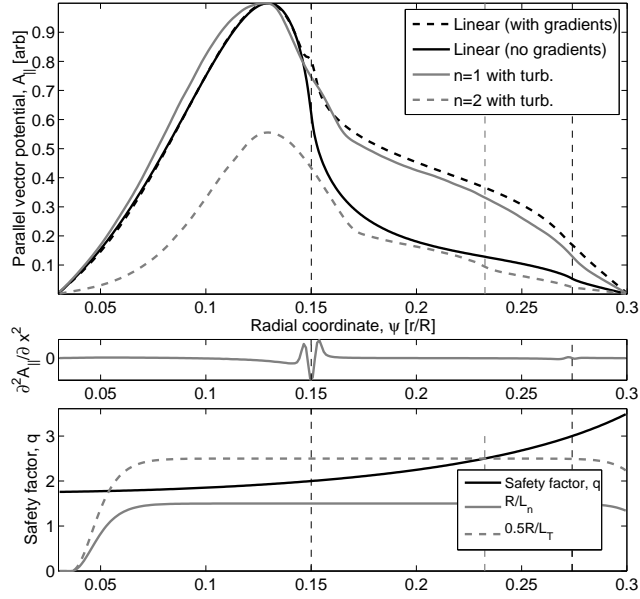}
\caption{(top) The radial profile of the parallel vector potential ($A_{||}$) for (black, solid) a linear calculation without the presence
of background density and temperature gradients, representative of the eigenfunction of the double tearing mode here with a $q=2$ and $q=3$ resonant layer.
(black, dashed) $A_{||}$ profile for a linear calculation with a background thermodynamic gradient $R/L_{n} = 1.5$, $R/L_{Te} = 5.0$. (grey, solid) $A_{||}$ profile in the
presence of electromagnetic turbulence once the island has been established.   
(middle) The second radial derivative of the linear eigenfunction, $A_{||}$ which shows the discontinuity at the resonant surfaces (bottom) the safety factor, q, profile (black, solid) used and the density ($R/L_{n}$, grey dashed) and temperature ($R/L_{T}$, grey solid) gradient  profiles.  Vertical dashed lines represent the positions of the resonant layers.}
\label{Eigenfunction}
\end{figure}

The gyrokinetic framework has been highly successful when applied to the numerical study of drift waves and turbulence, however the study of large scale instabilities
is in its infancy. Gyrokinetic calculations of the linear tearing mode have been performed but have
always concentrated on two dimensions \cite{Wan05,ROG07,NUM09}.  More recently, the kink instability
was studied using global gyrokinetic simulations \cite{Mish}.  The influence of the island on turbulence has been investigated in the presence of imposed island structures \cite{POL09,POL10,HorEPL}.  These studies have uncovered aspects of multi-scale behavior, including
modified zonal flow and vortex structures \cite{HorVor}.  Toroidal effects were shown to have a significant effect on temperature gradients \cite{Hor10}, bootstrap current \cite{HorVar,Berg09} and modify island growth rates \cite{IshWael,Mur11}. 

In this work we present fully self-consistent gyrokinetic calculations 
of a tearing mode in three dimensional toroidal geometry, with realistic plasma parameters in which 
turbulence, zonal flows and the magnetic island are allowed to evolve together. 

\begin{figure*}
\centering
\includegraphics[width=6.1cm,clip]{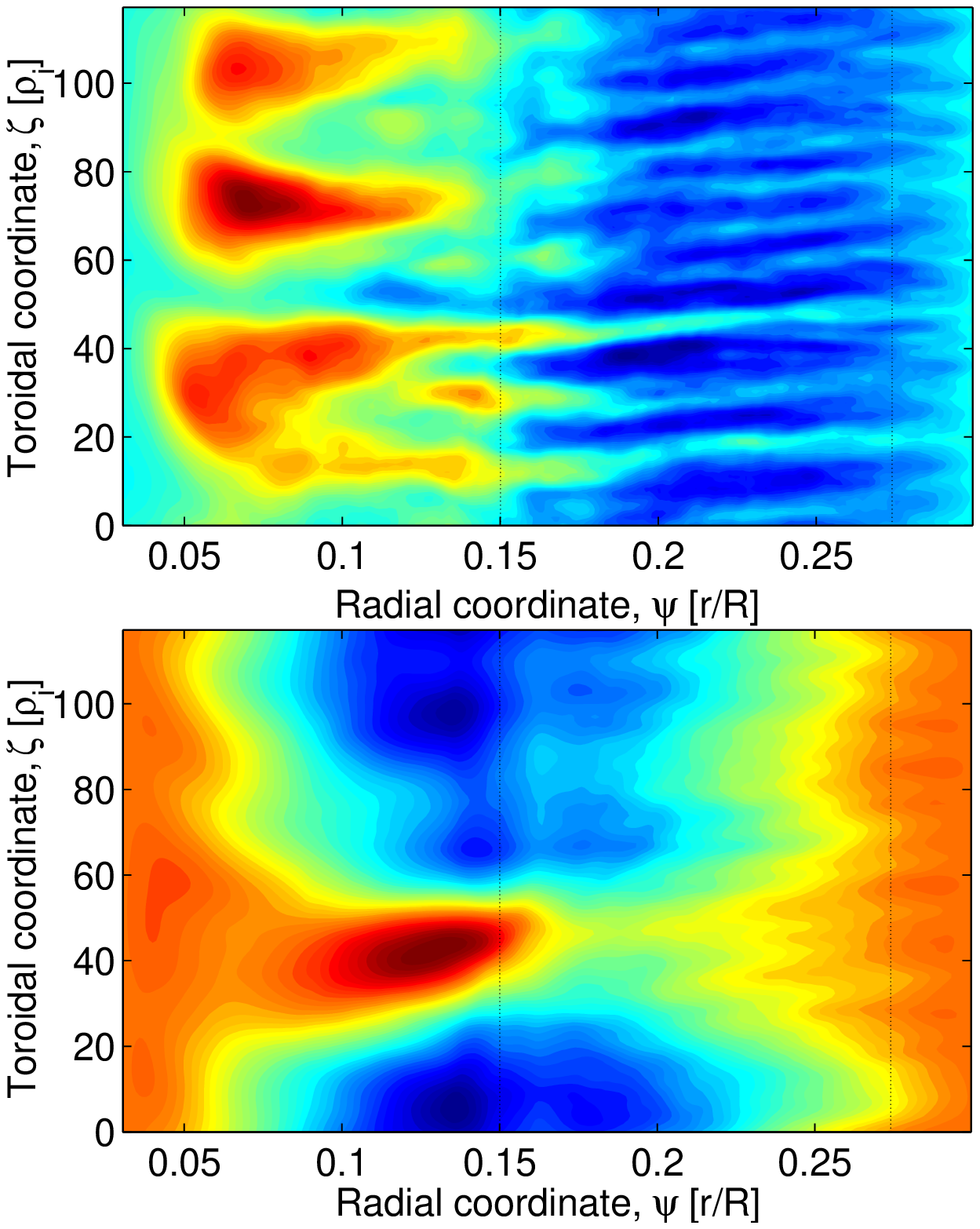}
\includegraphics[width=11.5cm,clip]{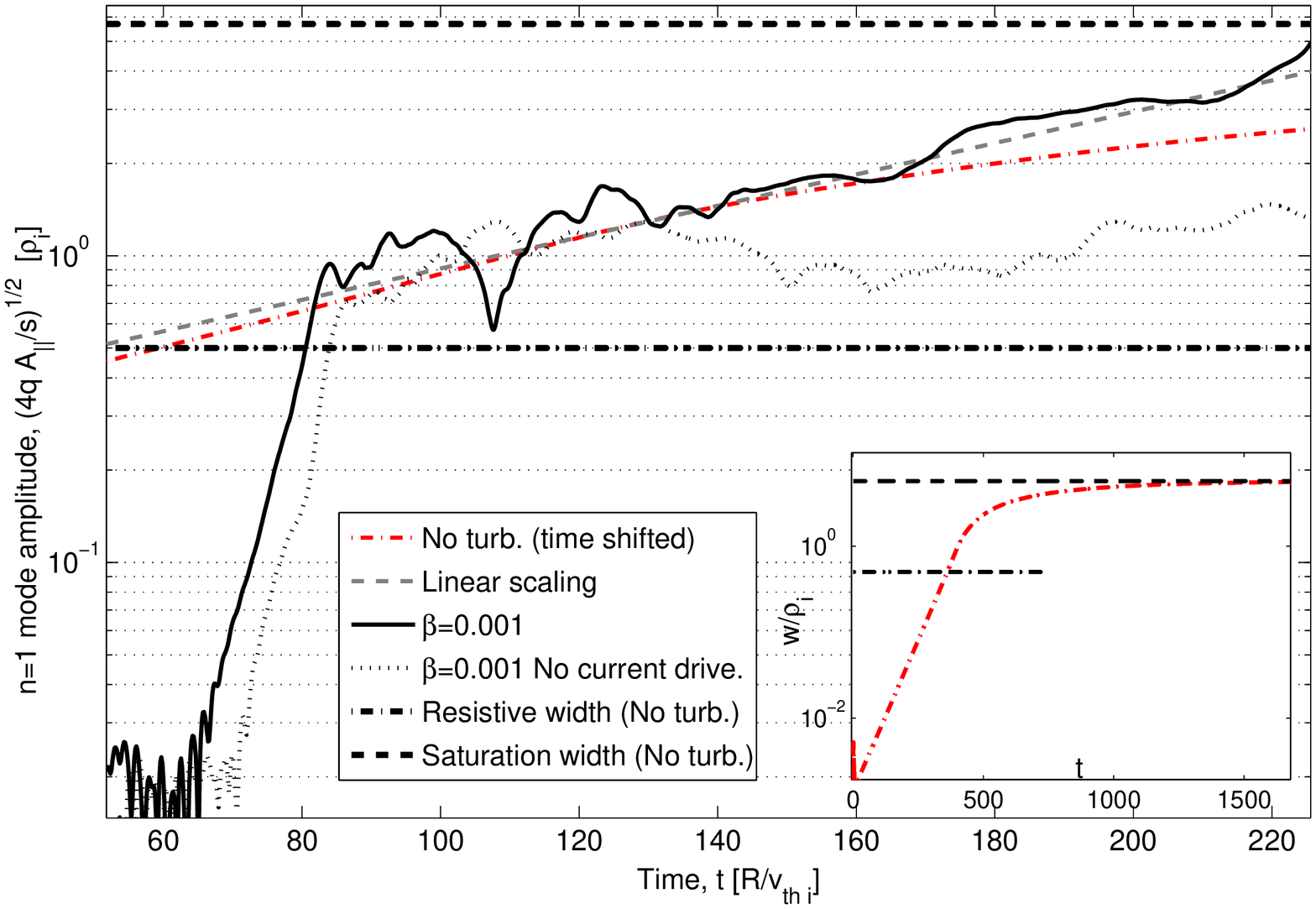}
\caption{(left, upper) A slice through the electrostatic potential and (lower) electromagnetic potential during a turbulence simulation at $t=165R/v_{th i}$ at a point
where the island structure has been established.  The vertical dashed lines represent the postitions of the resonant layers.  (right) The time trace of the width of the $m=2$, $n=1$ magnetic island (in units of $\rho_{i}$) in the presence of electromagnetic turbulence.  Plotted is the trace for two simulations, one without the tearing mode drive (dashed), and one with (black).  (grey dashed) represents the linear scaling.  The inlay shows the island width for a nonlinear simulation
without turbulence, showing the island evolve through the linear phase and nonlinear phase toward saturation.  The saturated width and resonant layer widths from this are plotted in both panels.}
\label{Islandtrace}
\end{figure*}

The global version of the gyrokinetic code, {\small GKW} is used, with details given in \cite{PEE09}.  The delta-$f$ approximation is employed, with the equation for the 
perturbed distribution function $f$, for each species, written in the form 
\be 
{\partial g \over \partial t} + (v_\parallel {\bf b} + {\bf v}_D) \cdot \nabla f +  {\bf v}_\chi\cdot \nabla g  
-{\mu B \over m}{{\bf B}\cdot \nabla B \over B^2}{\partial f \over \partial v_\parallel} = S, 
\label{gyrovlas}
\ee
where $S$ is the source term which is determined by the background distribution function, $F_{M}$, $\mu$ is the magnetic moment, $v_{||}$ is the velocity along the magnetic field, $B$ is the magnetic field strength, {\it m} and {\it Z} are the particle mass and charge number respectively. 
Here,  $g = f + (Ze/T)v_{\parallel}\langle A_{\parallel} \rangle F_{M}$ is used to absorb the time derivative 
of the parallel vector potential $\partial A_{\parallel}/\partial t$.  The thermal velocity $v_{\rm th}\equiv \sqrt{ 2 T_{ref} / m}$, and  the major radius of the magnetic axis ($R$) are used to normalise the length and time scales.  $T_{ref}$ is the temperature at $\psi=0.18$.
Here, $\rho_* = \rho_i / R$ is 
the normalised ion Larmor radius ($\rho_i = m_i v_{th} / e B$).  The velocities in Eq.~(\ref{gyrovlas}) are from left to right: the parallel motion along the 
unperturbed field ($v_\parallel {\bf b}$), the drift motion due to the inhomogeneous field 
(${\bf v}_D$), and the motion due to the perturbed electromagnetic field (${\bf v}_\chi$). 
Here, the angled brackets denote gyro-averaged quantities.  The gyro-average is calculated as a numerical average over a ring with a fixed radius equal to the 
Larmor radius, 
$\langle G \rangle ({\bf X}) = {1 \over 2 \pi }\oint {\rm d} \alpha \, G({\bf X} + \boldsymbol{ \rho})$
where $\alpha$ is the gyro-angle.  This gyro-average is used in both the evolution equation of the 
distribution function, as well as in the Poisson and Amp\'{e}re equations. The polarization in the former equation 
is linearized (i.e. is calculated using the Maxwell background rather than the full distribution function). A current profile is imposed on the electron background distribution which is calculated
self-consistently from the q-profile.  In this paper we use the model of Wesson et.~al. \cite{Wess}, where 
the current density profile is defined as $j=j_{0}(1+(r/a)^{2})$, which introduces an electron flow, $u_{e}$.  This enters 
the evolution equation via the source term, $\nabla F_{Me} = -2 \frac{v_{||}}{v_{th}^{2}} \nabla u_{e} F_{Me}$.    
The toroidal wave vectors are defined as, 
\begin{math}
k_\zeta^{I} \rho_i = 2 \pi n \rho_*,
\end{math}
where {\it n} is the toroidal mode number.  GKW uses a Fourier representation in the binormal direction, 
perpendicular to the magnetic field. The radial direction is treated using finite-differencing to include 
profiles in thermodynamic and geometry quantities.  The neoclassical term ($v_{D}\cdot\nabla F_{M}$) is neglected and thus the bootstrap current
drive is not present.

\begin{figure}
\centering
\includegraphics[width=8.0cm,clip]{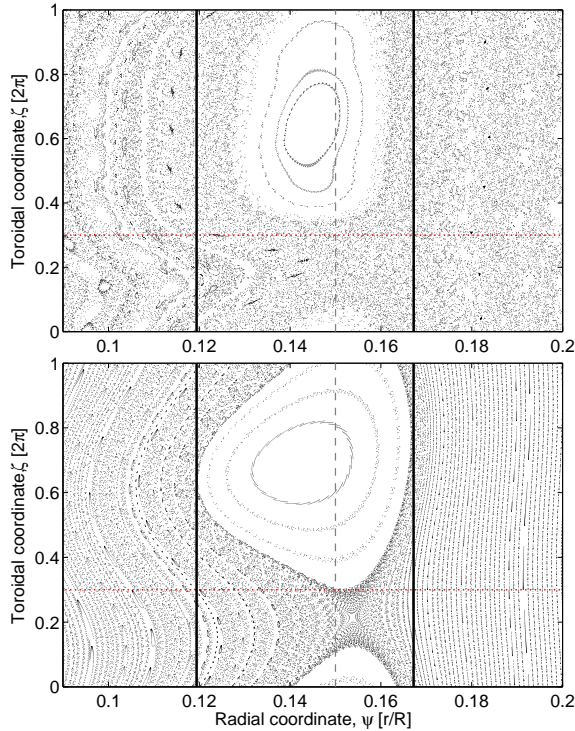}
\caption{Poincar\'{e} plot of the magnetic field lines as they pass through the low field side ($s=0$).  (Top) The field lines
trace the modified form of a magnetic island as it evolves in electromagnetic turbulence.  The red dots represent the original 
positions of the field lines.  For clarity the whole radial width of the domain is not shown. (Bottom) The equivalent plot
but with all higher mode numbers filtered out in post-processing, leaving only the $n=0$ and $n=1$ toroidal modes.  Vertical dashed lines denote the position
of the rational surface, and solid lines represent the positions of the island seperatrix when turbulence is neglected.}
\label{PoincareT}
\end{figure}

The simulations presented have the following parameters.  The concentric circular equilibrium geometry \cite{LAPcirc} is used.  The electron $\beta_{e}=10^{-3}$, aspect ratio, ${R}/{a}= 3$, Hydrogen mass ratio, $m_{i}/m_{e}= 1837$,  $q$ at the plasma edge, $q_{a} = 3.5$ with $q=2$, $q=2.5$ and $q=3$ rational surfaces ($q=m/n$, where m is the poloidal
mode number) in the computational domain, which, extends from the last closed flux surface to 10\% of the radius from the magnetic axis.  The normalised gyro-radius used is, $\rho_{*} =  5.10^{-3}$, which corresponds to a minimum  
mode number of, $k_{\zeta}\rho_{i} =  0.053$.  Used here, in total, were 36 toroidal modes giving a maximum $k_{\zeta}\rho_{i} =  1.86$. The ion and electron temperatures are
assumed to be equal, $T_{i} = T_{e}$.  Resolutions in the parallel, parallel velocity, magnetic moment, radial directions are, $N_{s}=64,N_{v}=64,N_{\mu}=16,N_{x}=256$ respectively.  
The radial resolution places three radial grid points within the singular layer.   In this paper we will concentrate on a single set of equilibrium temperature and density gradients, whose logarithmic scale lengths are $R/L_{T} = 5.5$ and $R/L_{n} = 1.5$ respectively.  The profiles are shown in bottom panel of Fig.~\ref{Eigenfunction}.

The current profile is found to provide a linearly unstable ($\Delta' > 0$) tearing mode with the familiar \cite{NISH98} linear
structure shown in Fig.~\ref{Eigenfunction}, showing the position of all the singular layers within the domain.  Shown in the 
middle panel is the second derivative of $A_{||}$ for the linear calculation clearly displaying the discontinuity in the solution at the rational surfaces.
The normalised collision frequency (to the trapping/detrapping rate), $\nu_{*} = 4\nu_{ei}/3\sqrt(\pi\epsilon^{3}) = 0.12$, where  the collision frequency is defined as
$\nu_{ei} = \frac{n_{i}e^{4}\log{\Lambda_{ei}}}{4\pi\epsilon_{0}^{2}m_{e}^{2}v_{e}^{3}}$, with these parameters the tearing mode is semi-collisional\cite{DRA77,FITZ10}.  In this paper we use a pitch-angle scattering of the electrons from the ions.

The inlay in Fig.~\ref{Islandtrace} shows the evolution of the island size in a nonlinear simulation in which only the $n=0$ and $n=1$ toroidal modes 
are kept.  In this case there is no small scale turbulence and initially, the mode grows exponentially until the island half-width ($w = \sqrt{4q A_{||}/B_{t}\hat{s}}$) 
equals the singular layer width (given by the dash-dotted line in both the inlay as well as the main figure).  At this island size the mode enters the Rutherford \cite{RUTH73} nonlinear
regime and grows algebraically until it finally saturates at an island
width $w = 5.7 \rho_i$ (indicated by the dashed lines).

When the electromagnetic turbulence is resolved, a perpendicular slice of which is shown in the left panel of Fig.~\ref{Islandtrace},
the mode displays a much more rapid growth in the initial phase (until approximately 100$R/v_{th i}$) compared with linear theory.  In this phase the mode amplitude ($n=1$, $k_{\theta}\rho_{i}=0.055$) closely traces
the amplitude of the turbulence ($A_{||}\propto |\phi|^{2}$) with $|\phi|^{2}$ taken from the peak of the turbulence spectrum ($k_{\theta}\rho_{i}=0.35$) implying a growth of the mode via nonlinear coupling with the electromagnetic ITG/TEM turbulence.  
The nonlinear coupling process produces $A_{||}$ structures 
approximately $\rho_{i}$ in size, larger than the singular layer width of linear theory.  Therefore, in the presence of turbulence linear tearing mode stability
is irrelevant since the turbulence, even at an electron beta $\beta_e= 0.1\%$, produces an island size for which linear theory is no longer
applicable.

Comparing the traces with (solid curve) and without (dotted curve) the tearing mode drive due to the radial gradient of the background
current, one observes that in the latter case the parallel vector potential ($A_{||}$) mode amplitude saturates after the initial phase, 
while in the former case it grows in amplitude throughout the simulation, developing increasing tearing parity and generating a growing magnetic island.    
From Fig.~\ref{Eigenfunction} which is taken from simulations where the drive is present, we see the distinctive radial eigenfunction of a $m=2$, $n=1$ tearing mode (linear growth rate, $\gamma=0.0235 v_{th i}/R$, calculated from GKW in linear mode).  
Therefore, even though electromagnetic turbulence is present the magnetic island continues to grow and maintain a coherent structure.  Turbulence does not disrupt the growth of the tearing mode, even for
island sizes of the order of the ion Larmor radius.

However, the shape of the magnetic island in a turbulent setting is modified as can be seen in Fig.~\ref{PoincareT} which 
shows a Poincar\'{e} plot of the magnetic field lines through the low field side, with the top figure retaining all the toroidal
modes in the magnetic field line tracing, while the bottom figure retains only the $n=0$ and $n=1$ mode (of the same nonlinear
simulation) in the field line trace.  The radial width of the closed island structure is significantly smaller when the small
scale modes are retained.  Even for the small $\beta_{e}$ that is used here, the island is 
about $30\%$ smaller, with the region around the separatrix and X-point highly stochastised.  The degree of stochastisation increases the higher the $\beta_{e}$, and is expected to have profound effects on the boundary layer around the island.
The polarization current stabilization \cite{WIL96}  connected with the rotation of the island that critically depends on the boundary
layer can be expected to be strongly modified.  The polarization current is, furthermore, highly dependent on the sign of the island rotation, which
when embedded in turbulence is seen to be in the ion diagmagnetic direction ($\omega=0.026 v_{th i}/R$).  In contrast, 
without turbulence, an island of the same size rotates
close to the electron diagmagnetic frequency ($\omega_{*e} = -(R/L_{n})(k_{th}\rho_{i})/2 = -0.038$).   In these simulations no evidence is seen for polarization current stabilization
of the mode, which, is invoked as a contributing factor in small islands.
 
Apart from the small scale structures that influence the island in a turbulent
setting large scales also develop.  Higher mode numbers, for 
example the $m=4$, $n=2$ mode (whose eigenfunction and corresponding resonant layers are shown in the dashed grey line 
in Fig.~\ref{Eigenfunction}), have a significant amplitude with respect to the $n=1$ mode and modify the island structure significantly. 
These higher modes as well as drift waves at this mode number are linearly stable and are generated here by nonlinear 
interactions.   Finally there is a slight shift in its radial position  of the island with respect to the $q=2$ 
rational surface of the background magnetic field (the latter is denoted by the vertical dashed black line).  This shift is caused by a finite amplitude 
of the vector potential in the $n=0$ toroidal mode.

Surprisingly after the seeding phase ($t > 80 R/v_{th i}$) it is seen that the island growth is closer to the linear rate (grey dashed line, Fig.~\ref{Islandtrace}), rather than the slower, non-linear rate (red dashed line).  
The latter curve is shown in full in the inlay and is calculated retaining only the $n=0$ and $n=1$ toroidal mode in the simulation, i.e. without
resolving the small scale turbulence.  Compared with the inlay, the curve representing the Rutherford growth in the main figure
has been shifted in time so that the amplitudes match the turbulence amplitude at the point where saturation is reached at approximately, $t=80 R/v_{th i}$.  The island at this point is much larger than the singular layer \cite{DRA77} 
width, which is the maximum island size at which linear theory is considered valid.  From Fig.~\ref{Eigenfunction} we see that
the $A_{||}$ profile from a time point ($t\sim180$) in a turbulence simulation is almost identical to its linear eigenfunction.  The fact that the island grows linearly at this amplitude implies that the turbulence
has modified the Rutherford nonlinear mechanism or its threshold.

\begin{acknowledgments}
A part of this work was carried out using the HELIOS supercomputer system at Computational Simulation Centre of International Fusion Energy Research Centre (IFERC-CSC), Aomori, Japan, under the Broader Approach collaboration 
between Euratom and Japan, implemented by Fusion for Energy and JAEA.
\end{acknowledgments}

\end{document}